\documentclass{elsart}
\usepackage{amssymb}
\usepackage{epsfig}
\usepackage{graphics}
\usepackage{color}
\definecolor{mygreen}{rgb}{0,.8,0}

\def\ifm#1{\relax\ifmmode#1\else$#1$\fi}
\def\DAF{DA\char8NE}  \def\sig{\ifm{\sigma}}
\def\f{\ifm{\phi}}   \def\epm{\ifm{e^+e^-}}
\def\ab{\ifm{\sim}}  \def\x{\ifm{\times}}
\def\gam{\ifm{\gamma}}  \def\pic{\ifm{\pi^+\pi^-}}
\def\pt#1,#2,{\ifm{#1\x10^{#2}}}  
\renewcommand{\to}{\ensuremath{\rightarrow}}

\def\up#1{$^{#1}$}    \def\L{\ifm{{\mathcal L}}}
\def\sig{\ifm{\sigma}}   
 \def\po{\ifm{\pi^0}}

\makeatletter
\newdimen\z@ \z@=0pt 
\newskip\z@skip \z@skip=0pt plus0pt minus0pt
\def\m@th{\mathsurround=\z@}
\def\ialign{\everycr{}\tabskip\z@skip\halign} 
\def\eqalign#1{\null\,\vcenter{\openup\jot\m@th
  \ialign{\strut\hfil$\displaystyle{##}$&$\displaystyle{{}##}$\hfil
      \crcr#1\crcr}}\,}
\makeatother

\newcommand{\aff}[2]{Dipartimento di Fisica dell'Universit\`a #1 e Sezione INFN, #2, Italy.}
\newcommand{\affd}[1]{Dipartimento di Fisica dell'Universit\`a e Sezione INFN, #1, Italy.}
\newcommand{\ie}{{\em i.e.}}

\begin{document}
\begin{frontmatter}

\title{
Study of the decay $\phi \rightarrow
\pi^+\pi^-\pi^0$ with the KLOE detector.  
}
\collab{The KLOE Collaboration}

\author[Na] {A.~Aloisio},
\author[Na]{F.~Ambrosino},
\author[Frascati]{A.~Antonelli},
\author[Frascati]{M.~Antonelli},
\author[Roma3]{C.~Bacci},
\author[Roma3]{F.~Bellini},
\author[Frascati]{G.~Bencivenni},
\author[Frascati]{S.~Bertolucci},
\author[Roma1]{C.~Bini}\footnote{Corresponding author: Cesare Bini, e-mail
  cesare.bini@roma1.infn.it, tel +390649914266, fax +39064957697},
\author[Frascati]{C.~Bloise},
\author[Roma1]{V.~Bocci},
\author[Frascati]{F.~Bossi},
\author[Roma3]{P.~Branchini},
\author[Moscow]{S.~A.~Bulychjov},
\author[Roma1]{G.~Cabibbo},
\author[Roma1]{R.~Caloi},
\author[Frascati]{P.~Campana},
\author[Frascati]{G.~Capon},
\author[Na]{T.~Capussela},
% ...

\address[Bari]{\affd{Bari}}
\address[Frascati]{Laboratori Nazionali di Frascati dell'INFN, 
Frascati, Italy.}
\address[Karlsruhe]{Institut f\"ur Experimentelle Kernphysik, 
Universit\"at Karlsruhe, Germany.}
\address[Lecce]{\affd{Lecce}}
\address[Na]{Dipartimento di Scienze Fisiche dell'Universit\`a 
``Federico II'' e Sezione INFN,
Napoli, Italy}
\address[Pisa]{\affd{Pisa}}
\address[Energ]{Dipartimento di Energetica dell'Universit\`a 
``La Sapienza'', Roma, Italy.}
\address[Roma1]{\aff{``La Sapienza''}{Roma}}
\address[Roma2]{\aff{``Tor Vergata''}{Roma}}
\address[Roma3]{\aff{``Roma Tre''}{Roma}}
\address[StonyBrook]{Physics Department, State University of New 
York at Stony Brook, USA.}
\address[Trieste]{\affd{Trieste}}
\address[Virginia]{Physics Department, University of Virginia, USA.}
\address[Beijing]{Permanent address: Institute of High Energy 
Physics, CAS,  Beijing, China.}
\address[Moscow]{Permanent address: Institute for Theoretical 
and Experimental Physics, Moscow, Russia.}
\begin{abstract}
We present a study of the reaction \epm\to\pic\po\ at the \f\ peak, $W=M(\f)$
=1019.4 MeV, observed 
with the KLOE detector at \DAF. The reaction is dominated by \f\ production and
decay, \epm\to\f\to\pic\po. 
From a fit to the Dalitz plot density distribution we obtain the $\rho$-meson
parameters for its three 
charge states. We also find the relative amplitudes for \f\to$\rho\pi$ and \f\to\pic\po\ and the cross section for $e^+e^-\rightarrow \omega\pi^0$ with $\omega\rightarrow\pi^+\pi^-$.
\end{abstract}
\end{frontmatter}

The decay of the $\phi$ meson to $\pi^+\pi^-\pi^0$, with a branching ratio
(BR) of \ab15.5\%, is dominated 
by the $\rho\pi$ intermediate states \cite{Parrour}
$\rho^+\pi^-$, $\rho^-\pi^+$, and $\rho^0\pi^0$ with equal amplitudes. We use
data 
from \epm\to\pic\po\ to determine the masses and widths of the three charge states
of the $\rho$-meson. CPT invariance requires equality of the masses and widths of
$\rho^+$ and $\rho^-$, while possible mass or width differences between
$\rho^0$ and $\rho^{\pm}$ are related to isospin-violating electromagnetic effects.  
Additional contributions to \epm\to\pic\po\ are the so called ``direct term'',
\f\to\pic\po \cite{vari}, 
and $\epm\to\omega\po$, $\omega\to\pic$.

We define the variables $x=T^+\!-T^-$ and $y=T^0$, where $T^{\;+\;,\;-,\;0}$
are the kinetic energies 
of the three pions in the center of 
mass system (CM). 
The Dalitz plot density distribution $D(x,y)$, in terms of the three amplitudes is given by:
\begin{equation}
\label{eqdalitz}
D(x,y)\propto|\vec p^{\;*}_{+}\times\vec p^{\;*}_{-}|^2|A_{\rho\pi}+A_{\rm dir}+A_{\omega\pi}|^2
\end{equation}
where $\vec p^{\;*}_{\pm}$ are the $\pi^{\pm}$ momenta in the CM and
$A_{\rho\pi}$, $A_{\rm dir}$ and 
$A_{\omega\pi}$ are the amplitudes described above.

Some 2 million $\pi^+\pi^-\pi^0$ events were collected with the KLOE detector
in $e^+e^-$ collisions at 
the Frascati \f-factory \DAF, in the fall of 2000. \DAF\ was run at a CM
energy $W=M(\f)$ and the 
integrated luminosity was \L=16 pb\up{-1}. In the following we present an
analysis of the Dalitz plot 
density distribution and new determinations of $\Gamma(\rho^{\;+-0})$,
$M(\rho^{\;+-0})$, 
$A_{\rho\pi}$, $A_{\rm dir}$, $A_{\omega\pi}$ and $\sig(\epm\to\omega\po,
\omega\to\pi^+\pi^-)$ at $W=1019.4$ MeV. 

The KLOE detector consists of 
a large volume drift chamber \cite{dch} (3.3 m length
and 2 m radius), operated with a 90\% helium-10\% isobutane gas mixture, and 
a sampling electromagnetic calorimeter \cite{calo} made of lead and 
scintillating fibres.
The whole detector is surrounded by a superconducting coil producing a solenoidal field $B$=0.52 T. The drift chamber momentum resolution
is $\sigma(p_\perp)/p_\perp$\ab 0.4\%. The calorimeter determines photon impact
points to an accuracy 
of 1 cm /$\sqrt{E({\rm GeV})}$ in the direction along the
fibres and of 1 cm in the transverse direction. Photon energies and arrival
times are measured with 
resolutions of $\sigma(E)/E=5.7\%/\sqrt{E({\rm GeV})}$ and $\sigma(t)=57{\rm
  ps}/\sqrt{E({\rm GeV})}\oplus 50$ ps. 
The solid angle coverage of the calorimeter is 98\% of 4$\pi$; however
particles from the 
interaction point with a polar angle with 
respect to the beam axis $\theta<15^\circ$ are shadowed by 
the low-$\beta$ insertion quadrupoles.
The trigger \cite{TRGnim} is based on the coincidence of at least 
two local energy deposits in the
calorimeter, above a threshold that ranges between 50 and 150 MeV. 
In order to reduce the trigger rate due to cosmic rays crossing 
the detector, 
events with a large energy release in the outermost
calorimeter planes are vetoed.

The \DAF\ beams collide with a crossing angle of 25 mrad, so the $\phi$ is
produced with a momentum in the horizontal plane $|\vec
  p_\phi|\sim$ 12.5 MeV.
$\pi^+\pi^-\pi^0$ events are selected by asking for two 
non-collinear tracks with opposite sign of curvature and polar angle $\theta>40^\circ$ which intersect the
interaction region. The acollinearity
cut ($\Delta\theta<175^\circ$) removes $e^+e^-\gamma$ events without incurring an
acceptance loss for the signal.
We then compute the missing mass,\\
$M_{\rm miss}= \sqrt{(E_\phi-E_{\pi^+}-E_{\pi^-})^2-|\vec p_\phi- \vec
  p_{\pi^+}-\vec p_{\pi^-}|^2}$ 
where $E$ and $\vec p$ are laboratory energies and momenta.
$M_{\rm miss}$ is required
to be within 20~MeV 
of the $\pi^0$ mass. This requirement corresponds to an effective energy cut of
$\leq20$ MeV on the total 
energy radiated because of initial state radiation (ISR).
\begin{figure}
\begin{center}
\mbox{\epsfig{file=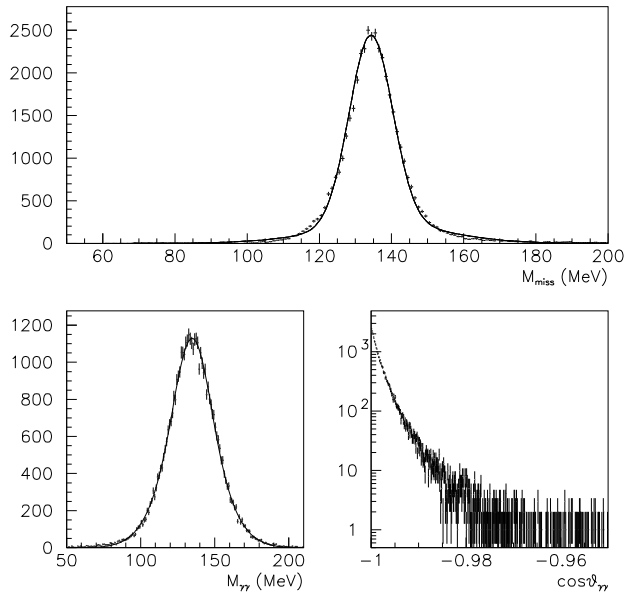,width=9cm}}
\end{center} 
\caption{Distributions of $M_{\rm miss}$ (top), $M_{\gam\gam}$, and
  $\cos\theta_{\gam\gam}$ 
(bottom left and right) for a sample of selected events. The rms widths of the
$M_{\rm miss}$ 
and $M_{\gam\gam}$ distributions are 5.5 MeV and 17 MeV, respectively. The solid lines are gaussian fits.}
\label{distributions}
\end{figure}
Two photons in the calorimeter are also required. A photon is defined as an
energy deposit larger than 
10 MeV with $21^\circ<\theta<159^\circ$ and an arrival time compatible with a particle
travelling 
at the speed of light, within 5$\sigma(t)$. The two-photon 
opening angle in the \po\ rest frame must satisfy $\cos\theta_{\gam\gam}<-0.98$. 

Fig. \ref{distributions} shows the distributions of the missing mass $M_{\rm
  miss}$, of the $\gam\gam$ 
invariant mass, and of cos$\theta_{\gamma\gamma}$ for a sample of selected
  events. 
Due to the large cross-section \footnote
{Here and in the following we consider visible cross-sections, 
not corrected for the effect of the radiative corrections.} 
for this final state with respect to other
  processes 
($\sigma_{\phi}\times {\rm BR} (\phi\to \pi^+\pi^-\pi^0)=460$ nb) and to the
  clean signature, 
the background to this process after the selection described is $\leq 10^{-5}$.
The Dalitz plot variables $x$ and $y$ are evaluated using the measured momenta
  of the charged pions, 
boosted to the center of mass system: $x=E_{+}^{*}-E_{-}^{*}$ and
  $y=E_\phi^{*}-E_{+}^{*}
-E_{-}^{*}-M_{\pi^0} =T_{\po}$.
$E_{\phi}$ and $\vec{p}_{\phi}$ 
are measured run by run using Bhabha scattering events. ISR lowers the mean
  \pic\po\ total energy by 
\ab130 keV. This value is used in the analysis with negligible effect on the results. 
The resolution on $x$ and $y$ is about 1~MeV over the full kinematical 
range.   

\begin{figure}
\begin{center}
\mbox{\epsfig{file=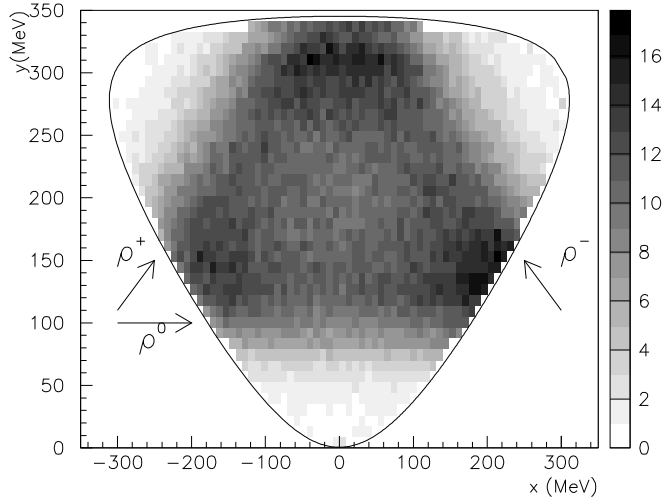,width=9cm}}
\end{center} 
\caption{Distribution of the number of events corrected for the efficiency and divided by
 $|\vec p^{\;*}_{+}\times\vec p^{\;*}_{-}|^2$. The grey scale is in arbitrary units.
 The plot contains 1.98 millions events in 1874 bins 8.75\x8.75 MeV\up2
each. Three broad bands corresponding to the three $\rho$ states are indicated.
The kinematical boundary is also shown.
}
\label{Dalitzdati}
\end{figure}

\begin{figure}
\begin{center}
\mbox{\epsfig{file=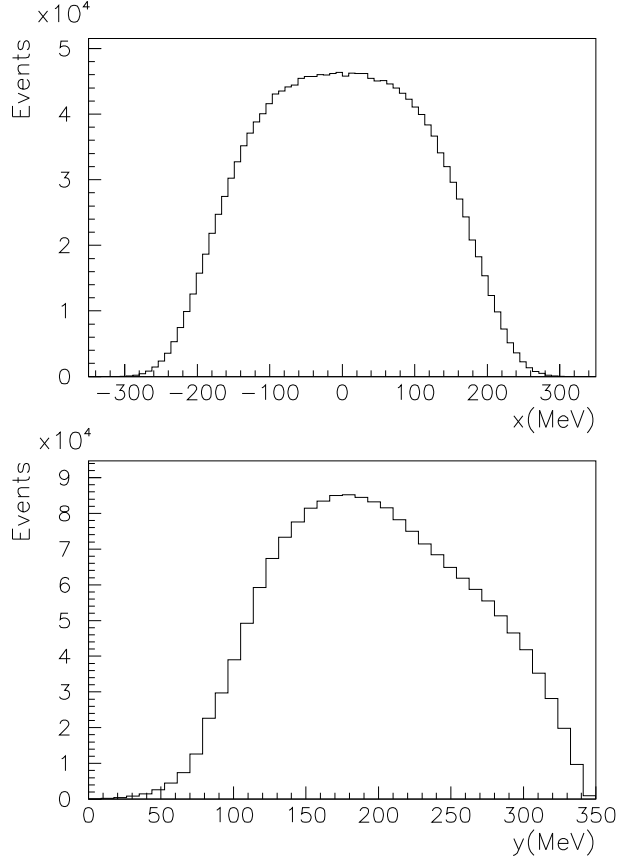,width=9cm}}
\end{center} 
\caption{Distributions in $x$ and $y$ of the number of events. 
The $\omega$ peak is not visible because of the smallness of the signal.}
\label{slopes}
\end{figure}

The Dalitz plot density distribution is shown in Fig. \ref{Dalitzdati}. 
In the plot the number of events corrected for the efficiency is shown 
divided by $|\vec p^{\;*}_{+}\times\vec
p^{\;*}_{-}|^2$. Three bands 
corresponding to
the three $\rho$ states are clearly evident. The
two-dimensional distribution 
is plotted in 8.75\x8.75 MeV\up2 bins. There are 1874 bins within the 
kinematic boundary. The bin
width is larger than the $x$ and $y$ resolution, but is small compared to 
the density variations of the 
Dalitz plot as can be seen in the $x$ and $y$ projections shown 
in Fig. \ref{slopes}.
Smearing effects due to the resolution are negligible. 

\begin{figure}
\begin{center}
\mbox{\epsfig{file=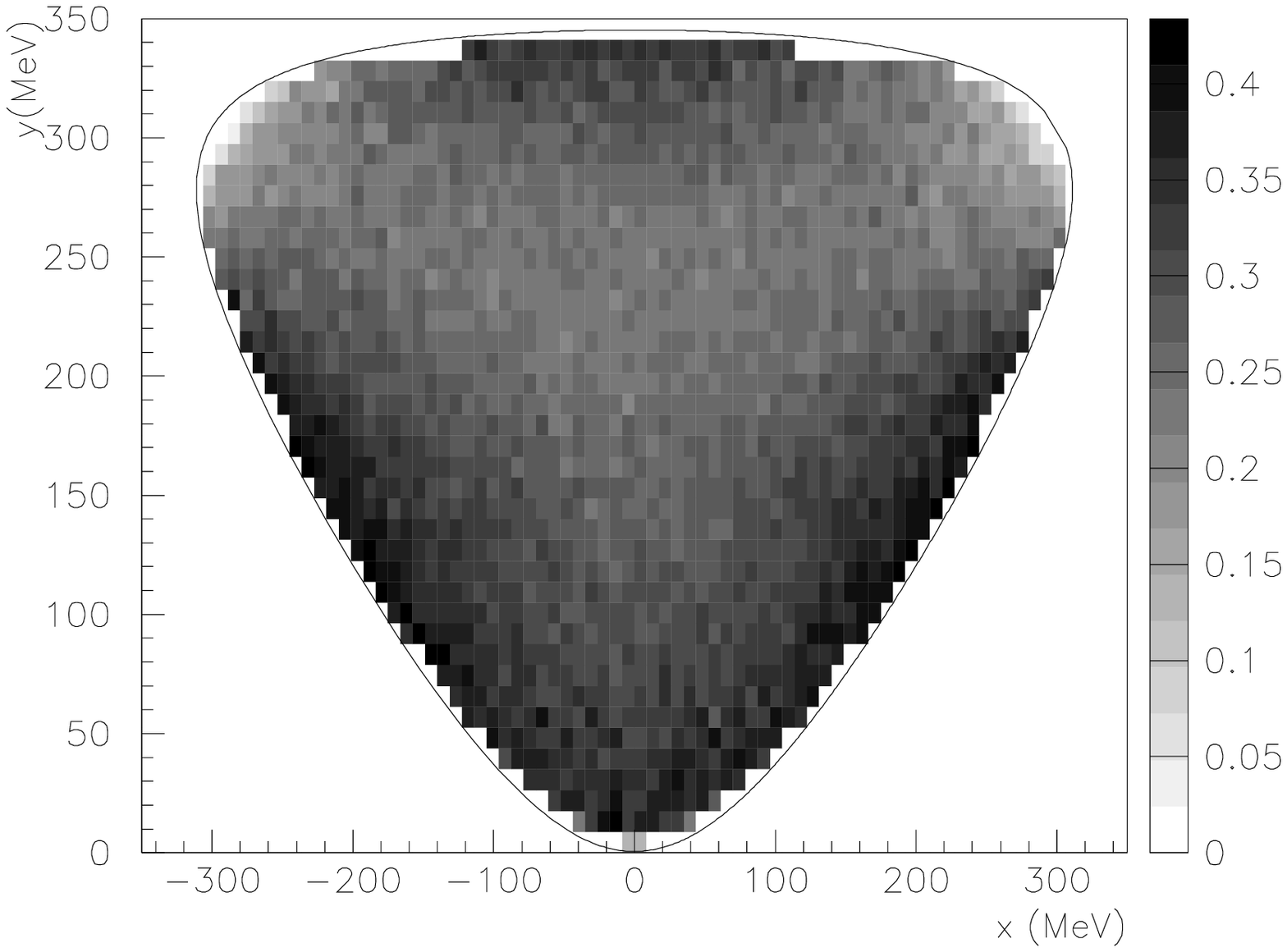,width=9cm}}
\end{center} 
\caption{Overall trigger and selection efficiency including the applied
  corrections as a function of the position in the Dalitz plot. 
}
\label{DalitzMC}
\end{figure}

Trigger and 
selection efficiencies have been evaluated as functions of $x$ and $y$.
A full Monte Carlo simulation of the detector has been used
with corrections based on control
samples of data. Corrections to the detection efficiency for low energy photons have been obtained 
using $e^+e^-\gamma$ events while analysis of tracking efficiency shows that the ratio of
data to Monte Carlo efficiency is very close to 1 and uniform in the relevant
momentum range. 
The trigger cosmic ray veto rejects \ab 5\% of the $\pi^+\pi^-\pi^0$
events. The amount and the distribution of the rejected events in the Dalitz
plot has been evaluated using downscaled samples of vetoed events. 
Fig. \ref{DalitzMC} shows the efficiency bin by bin 
obtained from a sample of $5\times 10^6$ Monte Carlo events. 
It can be noticed that the overall efficiency ranges between 20\% and 30\% 
within the Dalitz plot. The efficiency is dominated by acceptance cuts on
charged tracks and photons. In particular, the low efficiency in the upper corners is due 
to the fact that 
low momentum pions do not reach the drift chamber. The uncertainty on the efficiency 
is dominated by Monte Carlo statistics and
is on the order of 1\% in each bin.

The calibration of the momentum scale is checked using the measured missing
mass $M_{\rm miss}$. The central value of $M_{\rm miss}$ differs from $M_{\pi^0}$
by less than 0.1 MeV. 
The rms fluctuation of the missing mass across the entire
Dalitz plot is 0.25 MeV. 
The latter value is used in the estimates of systematic uncertainties for masses and widths.
    
The Dalitz plot distribution has been fitted according to
Eq. \ref{eqdalitz}.
The $\rho\pi$ term includes the line-shape of the three charge states of
the $\rho$ \cite{GouSak,Benayoun,KS} :
\begin{equation}
\label{eqrhopi}
A_{\rho\pi}=a_{\rho}\Sigma_{k}
{{M_{k}^2}\over {q_{k}^2-M_{k}^2+iq_{k}\Gamma_{(k)}(q_k^2)}}
\end{equation}
where $k=+,-,0$ is the $\rho$-meson charge and $\Gamma_{(k)}(q_k^2)$ is
defined as:
\begin{equation}
\label{eqgamma}
\Gamma_{(k)}(q_k^2)=\Gamma_{k}\left({{p_{\pi}(q_k^2)}\over
  {p_{\pi}(M_{k}^2)}}\right)^3\left({{M^2_{k}}\over {q_k^2}}\right)
\end{equation}
In the above, $q_k^2$ is the invariant mass of the appropriate pion pair, 
$\pi^i\pi^j$ with $i,j,k=+,-,0$; 
$M_k$ is the $\rho$ mass; $\Gamma_{k}$ the width; and $p_{\pi}$ is the
pion momentum in the $\rho$ center of mass.
The direct amplitude does not in general have the same phase as $A_{\rho\pi}$
and is taken as 
$A_{\rm dir}=a_de^{i\phi_d}$.
This term can be interpreted as the sum of $\rho$-like resonances of higher
mass like 
$\phi\rightarrow\rho'$(1415)$\pi$, which is not distinguishable from a 
constant term multiplied by 
$\times|\vec p_{+}^{*}\x\vec p_{-}^{*}|^2$. 
The $\omega\pi$ term is restricted to a horizontal band around $y=83.7$ MeV,
the kinetic energy 
of the \po\ in the $\omega\po$ final state.
The amplitude is given by:
\begin{equation}
\label{eqomegapi}
A_{\omega\pi}=a_{\omega}e^{i\phi_{\omega}}{{M_\omega^2}\over {q_{0}^2-M_\omega^2+iq_{0}\Gamma_{\omega}}}
\end{equation}
with the $\omega$ parameters fixed to PDG values \cite{PDG}. 
$a_{\rho}$, $a_{\omega}$ and $a_{d}$ are three dimensionless parameters.

\begin{figure}
\begin{center}
\mbox{\epsfig{file=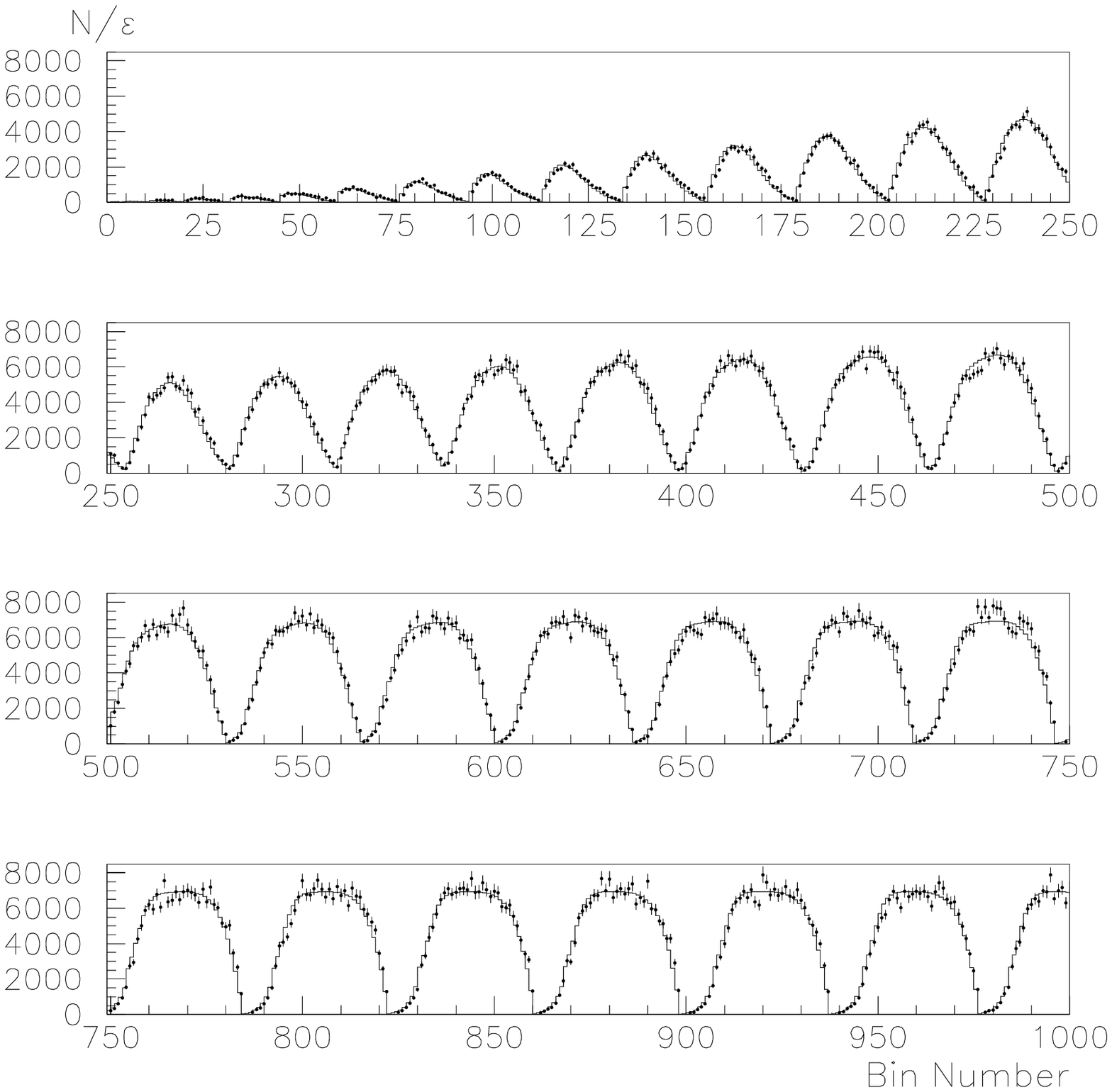,width=7.5cm}}
\vskip -0.85 cm
\mbox{\epsfig{file=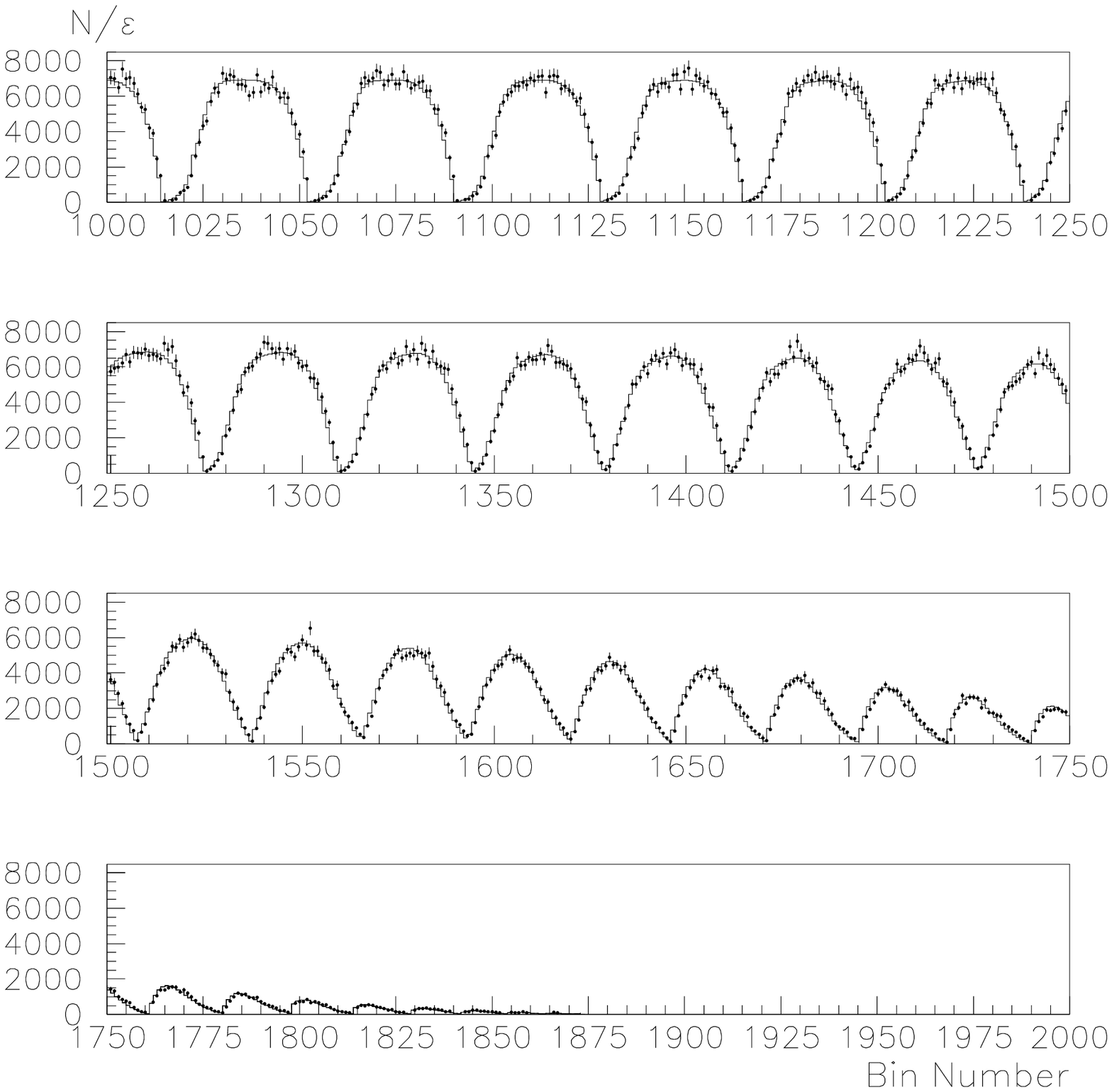,width=7.5cm}}
\end{center} 
\caption{Comparison between efficiency corrected data (points, $N/\epsilon$ is
  number of events per bin divided by the bin efficiency) 
and fitted function (histogram) 
  as a function of
  the bin number. The structure observed is due to the $y$ distributions
  for $x$ slices.
} 
\label{comparison}
\end{figure}

We perform a $\chi^2$ fit of the described density function to the data binned as described. The errors used are from data statistics and  Monte Carlo statistics for efficiency calculations. 
Three fits have been performed: a) a fit assuming CPT and isospin invariance, 
\ie\, $M_{\rho^0}=M_{\rho^+}=M_{\rho^-}$, $\Gamma_{\rho^0}=\Gamma_{\rho^+}=
\Gamma_{\rho^-}$; b) a fit assuming only CPT invariance \ie\ $M_{\rho^{+}}=
M_{\rho^-}$, $\Gamma_{\rho^+}=
\Gamma_{\rho^-}$; and finally c) without limitations on masses and widths. Moduli and phases of direct and $\omega\pi$ terms 
and $a_{\rho}$  are also free parameters.
There are 7 free parameters for fit a), 9 for fit b) and 11 for fit c). 
The number of degrees of freedom is 1874$-$7=1867 for fit a), 1865 for fit 
b) and 1863 for fit c). The results of the fits are shown in 
Tab. \ref{param} and a one-dimensional comparison of the data with the 
fit function (fit (c)) is shown in Fig. \ref{comparison}. 
\begin{table}
\caption{Fit results. Masses and widths are in MeV. The amplitudes are 
scaled in such a way that $a_{\rho}$=1. 
$\phi_d$ and $\phi_{\omega}$ are in radians. 
$\chi^2$ values and probabilities of the fits are given in the first row.
}
\vskip 0.1 in
\begin{center}
\begin{tabular}{|c|c|c|c|} \hline
 parameter & fit(a) & fit(b) & fit(c) \\ \hline
$\chi^2$ [$p(\chi^2)$]&1939 [12\%]&1914 [21\%]&1902 [26\%]\\ \hline          
$M_{\rho^0}$&                   &$775.9\pm0.5\pm0.5$&$775.9\pm0.6\pm0.5$\\ \cline{3-4}
$M_{\rho^+}$&$775.8\pm0.5\pm0.3$&$775.5\pm0.5\pm0.4$&$776.3\pm0.6\pm0.7$\\ \cline{4-4}
$M_{\rho^-}$&                   &                   &$774.8\pm0.6\pm0.4$\\ \hline
$\Gamma_{\rho^0}$&                   &$147.3\pm1.5\pm0.7$&$147.4\pm1.5\pm0.7$\\ \cline{3-4}
$\Gamma_{\rho^+}$&$143.9\pm1.3\pm1.1$&$143.7\pm1.3\pm1.2$&$144.7\pm1.4\pm1.2$\\ \cline{4-4}
$\Gamma_{\rho^-}$&                   &                   &$142.9\pm1.3\pm1.4$\\ \hline
 $a_d$           &$0.78\pm0.09\pm0.13$&$0.72\pm0.09\pm0.05$&$0.71\pm0.09\pm0.05$\\
 $\phi_d$        &$-2.47\pm0.08\pm0.08$&$-2.43\pm0.10\pm0.08$&$-2.43\pm0.10\pm0.08$\\ \hline
$a_{\omega}\times10^3$& $7.1\pm0.6\pm0.8$&$9.0\pm0.7\pm0.7$&$9.0\pm0.7\pm0.3$\\
$\phi_{\omega}$&$-0.22\pm0.11\pm0.04$&$-0.10\pm0.10\pm0.05$&$-0.10\pm0.10\pm0.07$\\
\hline
\end{tabular}
\vglue-.4cm
\end{center}
\label{param}
\end{table}

The uncertainties on the parameters are given as the sum of fit uncertainties 
(due to event statistics and efficiencies) and systematic uncertainties coming from two sources:
(a) the stability of the fit with respect to changes of the selection
cuts, and (b) the absolute momentum calibration.
Source (a) is dominant, and is evaluated by repeating the fit for data samples
obtained using different values of the cuts
and taking the rms variation of the parameters. Effects due 
to the modelling of ISR are automatically 
taken into account by the variation of the cut on the missing mass.

Fit (a) converges to an acceptable $\chi^2$ probability, showing that the
experimental distribution is consistent with CPT and isospin invariance.
Mass and width differences between states of charge obtained by fit (b) and
fit (c) are summarized 
in Tab. \ref{differ} where $M_{\rho^{\pm}}=(M_{\rho^+}+M_{\rho^-})/2$ and 
$\Gamma_{\rho^{\pm}}=(\Gamma_{\rho^+}+\Gamma_{\rho^-})/2$. 
The $\rho$ masses are significantly larger and the widths smaller than the PDG
averages 
($M_{\rho}=771.1\pm0.9$ MeV and $\Gamma_{\rho}=149.2\pm0.7$ MeV, see
\cite{PDG}) 
but are close to recent results \cite{Aleph,russihad,SND} for
the $\rho^0$ mass and width. 
    
\begin{table}[h]
\caption{Mass and width differences (in MeV) between $\rho$-mesons, from fit (b) and fit (c). Errors include the correlations between parameters.}
\vskip 0.1 in
\begin{center}
\begin{tabular}{|c|c|} \hline
 $M_{\rho^0}-M_{\rho^{\pm}}$ & $0.4\pm0.7\pm0.6$ \\
 $M_{\rho^+}-M_{\rho^-}$ & $1.5\pm0.8\pm0.7$ \\
 $\Gamma_{\rho^0}-\Gamma_{\rho^{\pm}}$ & $3.6\pm1.8\pm1.7$ \\
 $\Gamma_{\rho^+}-\Gamma_{\rho^-}$ & $1.8\pm2.0\pm0.5$ \\
\hline
\end{tabular}
\end{center}
\label{differ}
\end{table}
We observe significant contributions from the direct term and from the
$\omega\pi$ term: the coefficients $a_d$ and $a_{\omega}$ are significantly 
different from 0. If we define the weight of each contribution $\alpha$
($\alpha=\rho\pi$, ${\rm dir}$ and $\omega\pi$)
as \\ $I_{\alpha}=\int dxdy|A_{\alpha}|^2/\int
dxdy|A_{\rm tot}|^2$, where the amplitudes $A_{\alpha}$ are taken from the 
results of fit c) and $A_{\rm tot}$ is the sum of three amplitudes, 
we get the following weights:
\begin{equation}
I_{\rho\pi}=0.937
\end{equation}
\begin{equation}
I_{\rm dir}=8.5\times 10^{-3}
\end{equation}
\begin{equation}
I_{\omega\pi}=2.0\times 10^{-4}
\end{equation}
The sum of
the three weights is not equal to one, due to a sizeable interference term
between $A_{\rho\pi}$ and $A_{\rm dir}$ accounting for about 6\% of the total.


From $I_{\omega\pi}$, if we neglect interference, we obtain the visible cross section for the
non resonant process $e^+e^-\rightarrow\omega\pi^0$ with
$\omega\rightarrow\pi^+\pi^-$ at $W=1019.4$ MeV.  
\begin{equation}
\label{res2}
\sigma(e^+e^-\rightarrow\omega\pi^0\rightarrow\pi^+\pi^-\pi^0)=I_{\omega\pi}\times 
\sigma(e^+e^-\rightarrow\pi^+\pi^-\pi^0)= 92\pm15~{\rm pb}
\end{equation}
The ratio of this cross-section to the value $\sigma(e^+e^-\rightarrow\omega\pi^0
\rightarrow\pi^0\pi^0\gamma)=0.46\pm0.01\pm0.03~{\rm
  nb}$ obtained by KLOE in the 5 photons analysis \cite{KLOEGM} gives:
\begin{equation}
\label{res3}
R={{\rm BR}(\omega\rightarrow\pi^+\pi^-)\over
{\rm BR}(\omega\rightarrow\pi^0\gamma)}=0.20\pm0.04
\end{equation}
which compares well with the value 1.7/8.7=0.20$\pm$0.03 from \cite{PDG}.

\ack
We thank the DA$\Phi$NE team for their efforts in maintaining low
background running 
conditions and their collaboration during all
data-taking. We also thank G.F. Fortugno for 
his efforts in ensuring
good operations of the KLOE computing facilities. This work was
supported in part by DOE grant DE-FG-02-97ER41027; by 
EURODAPHNE, contract FMRX-CT98-0169; 
by the German Federal Ministry of Education and Research (BMBF) contract 
06-KA-957; 
by Graduiertenkolleg 'H.E. Phys.and Part. Astrophys.' of 
Deutsche Forschungsgemeinschaft, 
Contract No. GK 742;
by INTAS, contracts 96-624, 99-37; and by TARI, contract 
HPRI-CT-1999-00088.



\begin{thebibliography}{99}
\bibliographystyle{unsrt}

\bibitem{Parrour} A. London et al., Phys. Rev. {\bf 143} 1034 (1966);\\
G. Cosme et al., Phys. Lett. {\bf 48B} 155 (1974);\\
G. Parrour et al., Phys. Lett. {\bf B63} 357 (1976).
\bibitem{vari} S. Rudaz, Phys. Lett. {\bf B145} 281 (1984);\\
O. Kaymackalan et al., Phys.Rev. {\bf D30} 594 (1984);\\
T. Fujiwara et al., Progr. of Theor. Phys. {\bf73} 926 (1985).
\bibitem{dch} M. Adinolfi et al., KLOE coll., Nucl. Instr. and Meth. {\bf A488} 1 (2002).
\bibitem{calo} M. Adinolfi et al., KLOE coll., Nucl. Instr. and Meth. 
{\bf A482} 364 (2002).
\bibitem{TRGnim} M. Adinolfi et al., KLOE coll., Nucl. Instr. and Meth. {\bf
    A492} 134 (2002). 
\bibitem{GouSak} G. Gounaris J.J. Sakurai, Phys. Rev. Lett. {\bf21} 244 (1968).
\bibitem{Benayoun} M. Benayoun et al., Eur. Phys. J. {\bf C2} 269 (1998).
\bibitem{KS} J.H. K\"uhn and A. Santamaria, Zeit. Phys. {\bf C48} 445 (1990).
\bibitem{PDG} K. Hagiwara et al. (PDG), Phys.Rev. {\bf D66}, 010001 (2002).
\bibitem{Aleph} R. Barate et al., Zeit. Phys. {\bf C76} 15 (1997).
\bibitem{russihad} R.R. Akhmetshin et al., Phys. Lett. {\bf B527} 161 (2002).
\bibitem{SND} M.N. Achasov et al., Phys. Rev. {\bf D65} 03002 (2002).
\bibitem{KLOEGM} A. Aloisio et al., KLOE coll., Phys. Lett. {\bf B537} 21 (2002).
\end{thebibliography}
\end{document}